\documentclass{eptcs}

\usepackage{amsmath}
\usepackage{amssymb}
\usepackage{mathptmx}
\usepackage{url}
\usepackage{lmodern}
\usepackage[T1]{fontenc}
\usepackage{times}

\DeclareOldFontCommand {\oldstyle}{\usefont{U}{cmm}{m}{os}}%
                                  {\mathos}
\DeclareTextFontCommand{\textos}{\oldstylefamily}
\DeclareMathAlphabet   {\mathos}{U}{cmm}{m}{os}
\DeclareRobustCommand\oldstylefamily{%
            \not@math@alphabet\osshape\mathos
            \usefont{U}{cmm}\f@series{os}}

\newcount\clnb
\newcount\secnb
\newcount\ssecnb
\def\nsec#1{\clnb=0\ssecnb=0\global\advance\secnb by1\section{#1}}
\def\nssec#1{\global\advance\ssecnb by1\subsection{#1}}

\newcount\itemnb
\def\ITEM{\global\advance\itemnb by 1 {\bf(\the\itemnb)}\quad}

\long\def\kl#1#2#3{\global\advance\clnb by1
\par\smallbreak\noindent\ignorespaces\interlinepenalty=100
\edef\key{\noexpand\gdef\csname#1\endcsname{\the\secnb.\the\clnb}}\key
\textsc{#2}{\ \the\secnb.\the\clnb.} \enspace \textsl{#3}
\interlinepenalty=0 \medbreak\vskip0pt
\noindent\ignorespaces}

\newcount\fignb
\def\Topfig#1#2#3{\global\advance\fignb by1
\begin{figure*}[!t]
#1\hrulefill\quad \lower 2pt\hbox{\textbf{\textsf{Figure \the\fignb:
#3}}}\quad\hrulefill\\[2pt]
\edef\key{\noexpand\gdef\csname#2\endcsname{\the\fignb}}\key
\end{figure*} \noindent\ignorespaces}

\def\topfig#1#2#3{\global\advance\fignb by1
\begin{figure}[!t]
#1\hrulefill\quad \lower 2pt\hbox{\textbf{\textsf{Figure \the\fignb:
#3}}}\quad\hrulefill\\[2pt]
\edef\key{\noexpand\gdef\csname#2\endcsname{\the\fignb}}\key
\end{figure} \noindent\ignorespaces}
\def\nextfig{\global\advance\fignb by1{\the\fignb}\global\advance\fignb by-1}

\def\p{\kern -1pt{.}\kern 2pt}
\def\pp{\kern -1pt{.}\kern 0.7pt}

\def\LNCS{Lecture Notes in Comput.\ Sci.\ }

\def\kref#1#2#3#4#5#6{\bibitem{#1}
\textsc{#2, }\textsl{#3, }#4 ({\oldstyle #5}) #6.}

\def\krapp#1#2#3#4#5{\bibitem{#1}
\textsc{#2, }\textsl{#3, }#4 ({\oldstyle #5}).}
 
\def\setof#1#2{\{\,#1\mid#2\,\}}
\def\sset{\subseteq}
\def\impl{\Rightarrow}
\def\arr{\rightarrow}
\def\maps{\,\mathord{\mapsto}}
\def\substi#1#2{\{#1\maps#2\}}
\def\qpeq{\mathrel{::=}}
\def\dpeq{\mathrel{:=}}
\def\dom{\mathsf{dom}}
\def\equidef{\ \Leftrightarrow_{\rm def}\ }
\def\restric{\mathbin{\upharpoonright}}
\def\powerset#1{2^{#1}}

\def\mybar#1{
\setbox1=\hbox{$#1$}\dimen1=\ht1\dimen2=\wd1
\advance\dimen2 by -2pt
\hbox{\kern 1pt$\overline{\phantom{\hbox{\vrule height\dimen1 width\dimen2
depth 0pt}}}$\kern 1pt}\llap{$#1$}}

\def\emi{\hbox{\bf !}}
\def\Expr{\mathcal{E}\kern-1pt\mathit{xpr}}
\def\Val{\mathcal{V}\kern-1pt\mathit{al}}
\def\Var{\mathcal{V}\kern-1pt\mathit{ar}}
\def\Ref{\mathcal{R}\kern-1pt\mathit{ef}}
\def\Id{\mathcal{I}\kern-1pt\mathit{dent}}
\def\true{\mathit{tt}}
\def\false{\mathit{ff}}
\def\cond#1#2#3{(\mathsf{{if}}\ #1\ \mathsf{{then}}\ #2\ 
                 \mathsf{{else}}\ #3)}
\def\para{\mathop{\parallel}}
\def\rf{\mathsf{ref}}
\def\nil{(\!)}
\def\mo{b}
\def\L{\mathcal{L}}
\def\letin#1#2{(\mathsf{let}\ #1\ \mathsf{in}\ #2)}
\def\seq{\mathbin{\hbox{{\bf ;}}}}
\def\wr{\langle\mathsf{wr}\rangle}
\def\ww{\langle\mathsf{ww}\rangle}
\def\rr{\langle\mathsf{rr}\rangle}
\def\rw{\langle\mathsf{rw}\rangle}
\def\sync{\mathsf{sync}}
\def\lwsync{\mathsf{lwsync}}
\def\lw{\mathsf{lw}}
\def\trou{[]}
\def\EvCon{\mathbf{E}}
\def\Tid{\mathcal{T}\kern-1.5pt\mathit{id}}
\def\RC{\mathit{RC}}
\def\Dep{\mathcal{P}}
\def\ant{\mathrel{\blacktriangleleft}}
\def\reord#1{\mathrel{\Lsh^{#1}}}
\def\Mop{\mathcal{M}\kern-1pt\mathit{op}(\L)}
\def\tmp{\Sigma_{\L}}
\def\xarr#1#2{\xrightarrow[#1]{#2}}
\def\Visi{\mathcal{W}}
\def\mm{\mathcal{M}}
\def\branch#1#2#3{\cond{#1}{#2}{#3}}
\def\rlxt{\hookrightarrow}

\def\live{\mathsf{live}}
\def\rdt{\mathsf{rdt}}

\def\Sync{\mathcal{S}\kern-1pt\mathit{ync}}
\def\Bar{\mathcal{B}\kern-1pt\mathit{ar}}
\def\Read#1{\mathsf{rd}_{#1}}

\def\MRead#1{\mybar{\mathsf{rd}}_{#1}}

\def\Write#1{\mathsf{wr}_{#1}}

\def\Java{\textsf{J{\small AVA}}}
\long\def\OMIT#1{}

\begin{document}

\title{Relaxed Operational Semantics of Concurrent Programming Languages}

\author{G\'erard Boudol
\institute{INRIA -- Sophia Antipolis, France}
\and
Gustavo Petri
\institute{Purdue Univeristy, USA}
\and 
Bernard Serpette
\institute{INRIA -- Sophia Antipolis, France}
}
\def\titlerunning{Relaxed Operational Semantics of Concurrent Programming Languages}
\def\authorrunning{G. Boudol, G. Petri \& B. Serpette}
\maketitle

\begin{abstract}
We propose a novel, operational framework to formally describe the semantics of
concurrent programs running within the context of a relaxed memory model.
Our framework features a ``temporary store'' where the memory operations issued by the
threads are recorded, in program order. A memory model then specifies
the conditions under which a pending operation from this sequence is allowed to be 
globally performed, possibly out of order. The memory model also involves a
``write grain,'' accounting for architectures where a thread may read a write
that is not yet globally visible. Our formal model is supported by a software 
simulator, allowing us to run litmus tests in our semantics.
\end{abstract}

\nsec{Introduction}
The hardware evolution towards multicore architectures means that the most
significant future performance gains will rely on using concurrent programming 
techniques at the application level. This is currently supported by some general purpose
programming languages, such as \Java\ or C/C++.
The semantics that is assumed by the application programmer using such a concurrent language 
is the standard \textit{interleaving} semantics, also known as \textit{sequential consistency}
(SC, \cite{Lam}). This is also the semantics assumed by most verification methods.
However, it is well-known \cite{AdGha} that this semantics is \textit{not} the one
we observe when running concurrent programs in
optimizing execution environments, i.e.\ compilers and hardware architectures, which
are designed to run sequential programs as fast as possible. For 
instance, let us consider the program
\begin{equation}
\begin{array}{l}
p\dpeq\true\seq 
\\
r_0\dpeq\emi\,q
\end{array}
\ \big{\Vert}\
\begin{array}{l}
q\dpeq\true\seq
\\
r_1\dpeq\emi\,p
\end{array}
\label{tsoex}
\end{equation}
where we use ML's notation $\emi\,p$ for dereferencing the pointer -- or
reference, in ML's jargon -- $p$. 
If the initial state is such that the values of $p$ and $q$ are both $\false$, we cannot 
get, by the standard interleaving semantics, a final state where the value of both $r_0$
and $r_1$ is $\false$. Still, running this program may, on most  multiprocessor 
architectures, produce this outcome. This is the case for instance on a TSO machine 
\cite{AdGha} where the writes $p\dpeq\true$ and $q\dpeq\true$ are put in (distinct) 
buffers attached with the processors, and thus delayed with respect to the reads
$\emi\,q$ and $\emi\,p$ respectively, which get their value from the (not yet updated)
main memory. In effect, the reads are \textit{reordered} with respect to the
writes. Other reordering optimizations, which may also be introduced by compilers, yield 
similar failures of sequential consistency (see the survey \cite{AdGha}), yet sequential
consistency is generally considered as a suitable abstraction at the application
programming level.

Then a question is: how to ensure that concurrent programs running in a given optimized
execution environment appear, from the programmer's point of view, to be sequentially
consistent, behaving as in the interleaving semantics? 
A classical answer is: the program should not give rise to data races in its sequentially
consistent behavior, keeping apart some specific synchronization variables, like
locks. This is known as the ``DRF (Data Race Free) guarantee,'' that was first stated in
\cite{AdHia,Ghar}, and has been widely advocated since then (see \cite{AdBo,BoAd,JMM}). An
attractive feature of the DRF guarantee is that it allows the programmer to reason in
terms of the standard interleaving semantics alone. However, there are still some 
issues with this property. First, one would sometimes like to know what racy programs
do, for safety reasons as in \Java\ for instance, or for debugging purposes,
or else for the purpose of establishing the validity of program transformations 
in a relaxed memory model. 
Second, the DRF guarantee is more an axiom, or a contract, than a guarantee: once stated 
that racy programs have undefined semantics, how do we indeed 
guarantee that a particular implementation provides sequentially consistent semantics 
for race free programs?

Clearly, to address such a question, there is a preliminary problem to solve, namely: how
do we describe the actual behavior of concurrent programs running in a relaxed execution
environment? This is known to be a difficult problem. For instance, to the best of our 
knowledge, the \Java\ Memory Model (JMM) \cite{JMM} is still not sound.
Moreover, its current formal description is fairly complex. To our view,
this is true also regarding 
the formalization of the C++ primitives for concurrent programming \cite{BOSSW,BoAd},
or the formalization of the PowerPC memory model \cite{SarkPLDI}. 
Our intention here is \textit{not} to describe a specific 
memory model, be it a hardware, low-level one, or the memory model for a high-level
concurrent programming language, like \Java\ or C++. Our aim is rather to design 
a semantical framework that would be 
\\[-10pt]
\begin{itemize}
\item flexible enough to allow for the description
of a wide range of memory models;
\item simple enough to support the intuition of the programmer and the implementer;
\item precise enough to support formal analysis of programs.
\\[-10pt]
\end{itemize}
(Since we are talking about programs, there will be a programming language, but 
the particular choice we make is not essential to our work.)

To address the problem stated above, we adopt the \textit{operational} style advocated in
\cite{BouPeta,SewTSO}, which, besides being ``\textit{widely accessible to working
programmers}'' \cite{SewTSO}, allows us to use standard techniques to analyse and verify
programs, proving properties such as the DRF guarantee \cite{BouPeta} for instance. In
\cite{BouPeta,SewTSO}, write buffers are explicitly introduced in the semantic framework,
and their behavior accounts for some of the reorderings mentioned above.
The model we propose goes beyond the simple operational model for write buffering,
by introducing into the semantic framework a different intermediate structure, between the shared 
memory and the threads. The idea is to record in this structure the memory operations -- 
reads and write, or loads and stores, in low level terminology -- that are issued by the 
threads, in program order. We call the sequence of pending 
operations issued by the threads a \textit{temporary store}. 
Then these operations may be delayed, and finally performed, 
with regard to the global shared memory, out of order. To be globally performed, 
an operation from the temporary store must be allowed to overtake the operations that 
were previously issued, that is, the operations that precede it in the temporary store. 
Then a key ingredient in our model is the \textit{commutability predicate}, that 
characterizes, for a given memory model, the conditions under which an operation 
from the temporary store may be performed early. 
This accounts in particular for the usual relaxations of the program order, and
also for the semantics of synchronization constructs, like barriers.

In some relaxed memory models, some fairly complex behaviors arise that cannot be fully
explained by relaxations of the program order. These behaviors are caused by the failure of
write atomicity \cite{AdGha}. To deal with this feature, we introduce another key ingredient to 
characterize a memory model. In our framework, with each pending write is associated
a \textit{visibility}, that is the set of threads that can see it, and can therefore
read the written value. Depending on the memory model, and more specifically on the 
(abstract) communicating network topology between threads (or processors), not any set 
of threads is allowed to be a legitimate visibility. For instance, the Sequential 
Consistency model \cite{Lam} only allows the empty set, and the singletons to be visibility 
sets, meaning that only the thread issuing a write can see it before it is globally
performed. 
Then the definition of a memory model involves, besides the commutability predicate, 
a \textit{``write grain,''} which specifies which visibility a write is allowed to acquire.
This accounts for the fact that some threads can read others' writes early \cite{AdGha}. 
Our model then easily explains, in operational terms, the behavior of a series of ``litmus tests,'' 
such as IRIW, WRC, RWC and CC discussed in \cite{BoAd} for instance, and the tests 
from \cite{SarkPLDI}, designed to investigate the PowerPC architecture. Regarding this
particular memory model, we found only three cases where our formalization of 
the main PowerPC barriers is more strict than the one of \cite{SarkPLDI}. However,
these are cases where the behavior that our model forbids was never observed 
during the extensive experiments on real machines done by Sarkar \& al.\ (and reported
in files available on the web as a supplement to their paper). On the other hand, for 
all the litmus tests that can be expressed in our language, the behaviors that are observed 
in Sarkar's experiments on real machines are accounted for in our model, which therefore is 
not invalidated by these experimental results.
Needless to say, the experimental test suite provided by Sarkar \& al.\ was 
invaluable for us to see which behaviors the model should explain.
These litmus tests were, among others, run in a software simulator that
we have built to experiment with our semantics.

Compared to other formalizations of relaxed semantics, our model is truly operational.
By this we mean that it consists in a set of rules that specify what can be the next
step to perform, to go from one configuration to another. This contrasts with
\cite{BouPetb} for instance, where a whole sequence of steps is only deemed a valid behavior
if it can be shown equivalent to a computation in normal order. We notice that, again
constrasting \cite{BouPetb}, our model 
preserves a notion of \textit{causality}: a read can only return a value that is present in the
shared memory, or that is previously written by some thread.
Our notion of a temporary store is quite similar to the ``reorder box'' of
\cite{PaDi}, but formulated in the standard framework of programming language semantics.
In some approaches, including \cite{Atig} and \cite{BMS}, the various relaxations of the order 
of memory operations are described by means of rewrite rules on traces of memory operations
(which again are similar to our temporary store). Notice that permuting operations in a trace 
is, in general, a cyclic process.
Regarding the relaxation of write atomicity, and more specifically the read-others'-write-early 
capability (as illustrated by the IRIW litmus test in subsection \ref{earlysect} below,
where no relaxation of the program order is involved), the 
only work we know that proposes a formal operational formulation of this capability is 
\cite{SarkPLDI} which, to our view, provides a quite 
complicated semantics of this feature. We think that our formalization, by means of 
write visibility, is much simpler than the one of \cite{SarkPLDI}. Moreover, by relying 
on a concrete notion of state, our model should be more amenable to standard 
programming languages proof techniques, like for establishing 
that programs only exhibit sequentially consistent behavior \cite{BouPeta},
or more generally to achieve mathematical analysis and verification of programs. 
\\[5pt]
\textbf{\textsf{Note.}} The web page \url{http://www-sop.inria.fr/indes/MemoryModels/} contains 
a full version of the paper. The additional contents are explained in the text.

\nsec{The Core Language} 
Our language is a higher-order, imperative and concurrent language \`a la ML, that is
a call-by-value $\lambda$-calculus extended with constructs to deal with a mutable store. 
(This choice of a functional core language is largely a matter of taste.) 
In order to simplify some
technical developments, the syntax is given in administrative normal form.
In this way, only one construction, namely the application of a function to an argument,
is responsible for introducing an evaluation order (the program order).
Assuming given a set $\Var$ of variables, ranged over by $x,\ y,\ z\ldots$, the 
syntax is as follows:
\[\begin{array}{rcll}
v &\qpeq& x\ \mid\ \lambda xe
\ \mid\ \true\ \mid\ \false\ \mid\ \nil
&\quad\textsl{values}
\\[5pt]
\mo &\in& \Bar
&\quad\textsl{barriers}
\\[5pt]
e\in\L &\qpeq& v\ \mid\ (ve)\ \mid\ \cond{v}{e_0}{e_1}
&\quad\textsl{expressions}
\\[3pt]
&\mid& (\rf\,v)\ \mid\ (\emi\,v)
\ \mid\ (v_0\dpeq v_1)\ \mid\ \mo
\end{array}
\]
As usual, the variable $x$ is bound in an expression $\lambda xe$, and we consider
expressions up to $\alpha$-conversion, that is up to the renaming of bound variables.
The capture-avoiding substitution of a value $v$ for the free occurrences of $x$ in
$e$ is denoted $\substi{x}{v}e$.
We shall use some standard abbreviations like $\letin{x=e_0}{e_1}$
for $(\lambda xe_1e_0)$, which is also denoted $e_0\seq e_1$ whenever $x$
does not occur free in $e_1$.
We shall sometimes (in the examples) write expressions in standard syntax, which is
easily converted to administrative form, like for instance converting $(e_0e_1)$
into $\letin{f=e_0}{(fe_1)}$, or $(v\dpeq e)$ into $\letin{x=e}{(v\dpeq x)}$.

The barrier constructs are ``no-ops'' in the abstract (interleaving) semantics of the language. 
Such synchronization constructs are often considered low-level. However, we believe they can 
also be useful in a high-level
concurrent programming language, for ``relaxed memory aware'' programming (see \cite{BoAd}).
We do not focus on a particular set $\Bar$ here, so the language should actually be
$\L(\Bar)$, but in the following we shall give some examples of useful barriers, and 
see how to formalize their semantics.
In the full version of this paper we also consider constructs for spawning and joining
threads, and for locking references.

As usual, to formalize the operational semantics of the language, we have to extend it,
introducing some run-time values. Namely,
we assume given a set $\Ref$ of \textit{references}, ranged over by $p$, $q\ldots$.
These are the values returned by reference creation. In the examples we shall examine, 
the names $r_i$ suggest that such a reference should actually be regarded as a register, 
which is not shared with other threads.
We still use $e$ to range not only over expressions of the source language $\L$, but also
over expressions built with run-time values, that is, possibly involving references.

A step in the semantics consists in evaluating a redex inside an \textit{evaluation context}.
The syntax of the latter is as follows:
\[\begin{array}{rcll}
\EvCon &\qpeq& \trou\ \mid\ (v\,\EvCon)
&\quad\textsl{evaluation contexts}
\end{array}
\]
As usual, we denote by $\EvCon[e]$ the run-time expression obtained by filling the hole
in $\EvCon$ by $e$. 
The semantics is specified as small step transitions $C\arr C'$ between configurations
$C$, $C'$ of the form $(S,T)$ where $S$ and $T$ are respectively the \textit{store} and 
the \textit{thread system}. To define the latter, we assume given a set $\Tid$ of 
\textit{thread indentifiers}, ranged over by $t$. The store
$S$, also called here the \textit{memory}, is a mapping from a finite set $\dom(S)$ of 
references to values. The thread system $T$ is a mapping from a finite set $\dom(T)$ of 
thread identifiers, subset of $\Tid$, to run-time expressions. 
If $\dom(T)=\{t_1,\ldots,t_n\}$ and $T(t_i)= e_i$ we also write $T$ as

\Topfig{
\[\begin{array}{rcll}
(S,(t,\EvCon[(\lambda xev)])\para T) &\arr&
(S,(t,\EvCon[\substi{x}{v}e])\para T)
\\[3pt]
(S,(t,\EvCon[\cond{\true}{e_0}{e_1}])\para T) &\arr&
(S,(t,\EvCon[e_0])\para T)
\\[3pt]
(S,(t,\EvCon[\cond{\false}{e_0}{e_1}])\para T) &\arr&
(S,(t,\EvCon[e_1])\para T)
\\[3pt]
(S,(t,\EvCon[(\rf\,v)])\para T) &\arr&
(S\cup\{p\mapsto v\},(t,\EvCon[p])\para T)
&\quad{\rm if}\ p\not\in\dom(S)
\\[3pt]
(S,(t,\EvCon[(\emi\,p)])\para T) &\arr&
(S,(t,\EvCon[v])\para T)
&\quad{\rm if}\ S(p)=v
\\[3pt]
(S,(t,\EvCon[(p\dpeq v)])\para T) &\arr&
(S[p\dpeq v],( t,\EvCon[\nil])\para T)
\\[3pt]
(S,(t,\EvCon[\mo])\para T) &\arr&
(S,(t,\EvCon[\nil])\para T)
\end{array}
\\[-2pt]
\]
}{sourceopsemfig}{Reference Operational Semantics}
\[
(t_1, e_1) \para \cdots \para (t_n, e_n)
\]
The reference operational semantics, that is the standard interleaving semantics, is 
given in Figure~\sourceopsemfig.

\nsec{Relaxed Computations}\label{RelaxSemSection}
\nssec{Preliminary Definitions}
The relaxed operational semantics is formalized by means of small steps transitions
\[
\RC\xarr{\mm}{}\RC'
\]
between relaxed configurations $\RC$ and $\RC'$. The $\mm$ parameter is the 
\textit{memory model}. 
Let us first describe the relaxed configurations. For this we need to introduce
some technical ingredients.
In the relaxed semantics a read can be issued by a thread, evaluating a subexpression
$(\emi\,p)$, while not immediately returning a value. In this way the read can be
overtaken by a subsequent operation. To model this, we shall dynamically assign to each
read operation a unique identifier, returned as the value read. That is, we extend the
language with names, or \textit{identifiers}, to point to future values. 
The set $\Id$ of identifiers is assumed to be disjoint from
$\Var\cup\Ref$, and is ranged over by $\iota$. We shall use $\varrho$ to
range over $\Ref\cup\Id$.  The identifiers $\iota\in\Id$ are \textit{values} in the extended
language, still denoted by $v$, but notice that $\Val$ denotes the set of (not relaxed)
values, that do not contain any identifier $\iota$. We shall require that only true values,
not relaxed ones, can be stored. It should be clear that substituting a relaxed value $v$
for an identifier $\iota$ in an expression $e$ results in a valid expression, denoted
$\substi{\iota}{v}e$.

Our next technical ingredient is the set $\Mop$ of \textit{memory operations}
in the language $\L$. These
represent the instructions that are issued by the threads, but are not necessarily 
immediately performed. The set $\Mop$ of memory operations comprises the barriers 
$\mo\in\Bar$ and the \textit{read} and 
\textit{write} operations, respectively denoted $\Read{\varrho,\iota}$ and 
$\Write{\varrho,v}^{W,I}$ where $\varrho\in\Ref\cup\Id$, $\iota\in\Id$, $W\sset\Tid$
is a set of thread names, and $I\sset\Id$ is a set of identifiers.
We call the set $W$ in $\Write{\varrho,v}^{W,I}$ the 
\textit{visibility} of the write (we comment on this, and on the $I$ component, below). 
Finally, we introduce operations of the form $\MRead{\iota}$ that we call 
a \textit{read mark}, meaning that a read has occurred, where $\iota$ serves as identifying the 
corresponding write.
That is, the syntax of memory operations is as follows:
\[
\xi\in\Mop\ \qpeq\ \Read{\varrho,\iota}\ \mid\ \MRead{\iota}
\ \mid\ \Write{\varrho,v}^{W,I}\ \mid\ \mo
\]
We can now define a \textit{relaxed configuration} $\RC$ as a triple $\RC=(S,\sigma,T)$
where $S$ and $T$ are as above, and
$\sigma$ is a sequence of pairs $(t,\xi)$, where $t\in\Tid$ is a thread name
and $\xi\in\Mop$ a memory operation.
The meaning of $(t,\xi)$ in a sequence $\sigma$ is that $\xi$ is a memory operation 
issued by thread $t$. The sequence $\sigma$ then records the pending memory
operations issued by the threads, which will not necessarily be performed (on the shared
memory) in the order in which they appear in $\sigma$. We shall call such a $\sigma$ a
\textit{temporary store}. We denote by $\tmp$  the set $\Tid\times\Mop$,
so that the set of temporary stores is $\tmp^*$, the set of finite sequences over
$\tmp$. We denote by $\varepsilon$ the {\rm empty sequence}, and we write 
$\sigma\cdot\sigma'$ for the {\rm concatenation} of the two sequences $\sigma$ and 
$\sigma'$. We say that a relaxed configuration $(S,\sigma,T)$ is \textit{normal} 
whenever $\sigma=\varepsilon$, and no expression occurring in the configuration (that is,
in the store $S$ or the thread pool $T$) contains an identifier. 

\nssec{The Relaxed Semantics}
We present the relaxed semantics in two parts: the first one describes the evaluation of
the threads, that is, the contribution of the $T$ component in the semantics, and the
second one explains how the memory operations from the temporary store $\sigma$ are
performed. 
One could say that the instructions executed by the threads are ``locally
performed,'' while the operations executed from the temporary store will be 
``globally performed,'' as their effect is made visible to the other threads.
The particular memory model $\mm$ is irrelevant to the local evaluation of threads,
\global\advance\fignb by1
and therefore in Figure~\the\fignb, which presents this evaluation, we simplify
\global\advance\fignb by-1
$\xarr{\mm}{}$ into $\rlxt$. In the rules for reducing $(\rf\,v)$ and $(\emi\,\varrho)$,
``$p$ fresh'' and ``$\iota$ fresh'' mean that $p$ and $\iota$ do not occur in the
configuration.

\Topfig{
\[\begin{array}{rcll}
(S,\sigma,(t,\EvCon[(\lambda xev)])\para T) 
&\rlxt&
(S,\sigma,({t},\EvCon[\substi{x}{v}e])\para T)
\\[7pt]
(S,\sigma,(t,\EvCon[\branch{\true}{e_0}{e_1}])\para T) 
&\rlxt&
(S,\sigma,({t},\EvCon[e_0])\para T)
\\[7pt]
(S,\sigma,(t,\EvCon[\branch{\false}{e_0}{e_1}])\para T) 
&\rlxt&
(S,\sigma,({t},\EvCon[e_1])\para T)
\\[7pt]
(S,\sigma,(t,\EvCon[(\rf\,v)])\para T) 
&\rlxt&
(S,\sigma\cdot(t,\Write{p,v}^{\emptyset,\emptyset}),({t},\EvCon[p])\para T)
&\quad p\ {\rm fresh}
\\[5pt]
(S,\sigma,(t,\EvCon[(\emi\,\varrho)])\para T) 
&\rlxt&
(S,\sigma\cdot(t,\Read{\varrho,\iota}),
({t},\EvCon[\iota])\para T)
&\quad\iota\ {\rm fresh}
\\[5pt]
(S,\sigma,(t,\EvCon[(\varrho\dpeq v)])\para T)
&\rlxt&
(S,\sigma\cdot(t,\Write{\varrho,v}^{\emptyset,\emptyset}),
({t},\EvCon[\nil])\para T)
\\[7pt]
(S,\sigma,(t,\EvCon[b])\para T) &\rlxt&
(S,\sigma\cdot(t,b),({t},\EvCon[\nil])\para T)
\end{array}
\\[-2pt]
\]
}{relaxopsemfigT}{$\mm$-Relaxed Operational Semantics (Threads)}

The relaxed semantics differs from the reference semantics in several ways. The main
difference is that the effect on the memory -- if any -- of evaluating the code is
delayed. 
Namely, instead of updating the memory, the effect of evaluating $(p\dpeq v)$, or
more generally $(\varrho\dpeq v)$ where the exact reference to update may still be
undetermined, consists in recording the write operation, with a default empty visibility, 
at the end of the sequence of pending memory operations. Creating a reference, reducing
$(\rf\,v)$, has the same effect, once
a new reference name is obtained. Reducing a dereferencing operation $(\emi\,\varrho)$ 
does not immediately return a proper value, but creates and returns a fresh identifier 
$\iota\in\Id$, to be later bound to a definite value, while appending a 
corresponding read operation to the temporary store. A barrier just appends itself at 
the end of the temporary store. Notice that the rules of Figure~\relaxopsemfigT\ are 
not concerned with the store $S$. 
As an example, considering the thread system $T$ of
Example (\ref{tsoex}) given in the Introduction, assigning the thread names
$t_0$ to the thread on the left and $t_1$ to the one on the right, assuming $\iota_0$ 
and $\iota_1$ to be fresh identifiers, and executing $t_0$ 
followed by $t_1$ we can reach the following temporal store:
\[
\sigma = (t_0,\Write{p,\true}^{\emptyset,\emptyset})\cdot(t_0,\Read{q,\iota_0})\cdot
(t_1,\Write{q,\true}^{\emptyset,\emptyset})\cdot(t_1,\Read{p,\iota_1})
\]
A relaxed configuration $(S,\sigma,T)$ can also perform actions that originate
from the temporary store $\sigma$. These steps are performed independently from
the evaluation of threads, in an asynchronous way. To define these transitions, we need to
say a bit more about the memory model $\mm$. We shall not focus here on a particular 
memory model, since our purpose is to design a general framework for describing the
semantics of concurrent programs in a relaxed setting. 
However, we shall make some minimal hypotheses about the $\mm$ parameter.
But let us first say what $\mm$ consists of.
We assume that this is a pair $\mm=(\reord{},\Visi)$ made of a \textit{commutability} 
predicate $\reord{}$ and a ``\textit{write grain}'' $\Visi$.
These two components provide a formalization of the approach of Adve and
Gharachroloo in \cite{AdGha}, who distinguish these two key features as the basis for
categorizing memory models.

The commutability predicate delineates the relaxations of the program order that are allowed 
in the weak semantics under consideration, and in particular it provides semantics for barriers. 
This first component $\reord{}$ of a memory model is a subset of $\tmp^*\times\tmp$, that
is a binary predicate relating temporary stores $\sigma\in\tmp^*$ with issued operations
$(t,\xi)\in\tmp$. This predicate is expressing which operations
issued by some thread are allowed to be performed early, that is, out of order in the 
relaxed semantics. Indeed, if the temporary store is $\sigma\cdot(t,\xi)\cdot\sigma'$
with $\sigma\reord{}(t,\xi)$, then the operation $\xi$ from thread $t$ may, in general, 
be globally
performed, as if it were the first one, and removed from the temporary store. We read
$\sigma\reord{}(t,\xi)$ as: $(t,\xi)$ may overtake $\sigma$, or: $\sigma$ allows
$(t,\xi)$ to be performed. We assume, as an axiom satisfied by any memory model, 
that the first operation in the temporary store is always allowed to execute, that is,
for any $\xi$ and $t$:
\[
\varepsilon\reord{}(t,\xi)
\tag{E}
\]
The $\Visi$ component of a memory model
is a set of subsets of $\Tid$, comprising the set of the allowed write 
visibilities. In the relaxed semantics, with each write operation $\Write{\varrho,v}^{W,I}$ 
is associated a visibility $W\kern-2pt$, which is a (possibly empty) set of thread identifiers. 
(We delay the discussion of the set $I$ to the subsection~\ref{mmrequirements}.)
The default visibility of a write when it is issued, as prescribed in Figure~\relaxopsemfigT, 
is $\emptyset$, so we assume that for any memory model this is an allowed visibility, that 
is $\emptyset\in\Visi$.
The visibility of a write may dynamically evolve (within $\Visi$), but we shall
assume that it can only grow. The threads in $W$ see the write, while in the temporary
store, and these threads can therefore read the corresponding value, possibly before
it is globally visible (in that case the $I$ component of the write is extended).
The $\Visi$ component allows us to deal with \textit{write atomicity}, or, more
generally, with the extent to which the threads are allowed to read each others writes. 
In a hardware architecture, this is
determined by a particular topology and behavior of the interconnection network.
Thus, for example, assuming three different threads $t$, $t'$ and $t''$, a write 
$\Write{\varrho,v}^{\{t,t'\},I}$ in the temporal store can be prematurely 
read by thread $t$ and $t'$ but not from thread $t''$.

We can now formulate the rules for the $\xarr{\mm}{}$ transitions as regards the 
memory. 
\global\advance\fignb by1
These are given in Figure~\the\fignb, with $(\reord{},\Visi)=\mm$. 
\global\advance\fignb by-1
In the rule $R2$ we use a restricted commutability predicate $\sigma\reord{\Bar}(t,\xi)$, 
ignoring the operations from $\sigma$ that are not synchronization operations, that is:
\[
\sigma\reord{\Bar}(t,\xi)\ \equidef\ \sigma\restric\Bar\reord{}(t,\xi)
\]
where $\sigma\restric\Bar$ is the restriction of the sequence $\sigma$ to the 
set $\Bar$, that is the subsequence of $\sigma$ containing only the issued barriers.

\def\rlx#1{\xarr{\reord{},\Visi}{#1}}
\def\rlxm{\xarr{\reord{},\Visi}{}}
\Topfig{
\[\begin{array}{rcll}
(S,\sigma, T) &\rlxm&
(S,\substi{\iota}{v}(\sigma_0\cdot\sigma_1, T))
&\kern21pt R1\ (\textit{read})
\\[3pt]&&
{\rm if}\quad \sigma = \sigma_0\cdot(t,\Read{p,\iota})\cdot\sigma_1\ \&\
\sigma_0\reord{}({t},\Read{p,\iota})\ \&\ S(p)=v\ 
\\[3pt]
(S,\sigma, T) &\rlxm&
(S,\substi{\iota}{v}(\sigma_0\cdot(t',\Write{p,v}^{W,I\cup\{\iota\}})\cdot\sigma_1
\cdot(t,\MRead{\iota})\cdot\sigma_2, T))
&\kern-1pt R2\ (\textit{read early})
\\[3pt]&&
{\rm if}\quad \sigma = \sigma_0\cdot(t',\Write{p,v}^{W,I})\cdot\sigma_1\cdot(t,\Read{p,\iota})
\cdot\sigma_2\ \&\ {t}\in W\ \&\ 
\\[3pt]&&
\phantom{{\rm if}\quad }
\sigma_1\reord{}({t},\Read{p,\iota})\ \&\ \sigma_0\reord{\Bar}({t},\Read{p,\iota})
\\[3pt]
(S,\sigma, T) &\rlxm& (S,\sigma_0\cdot\sigma_1, T)
&\kern21pt R3\ (\textit{read})
\\[3pt]&&
{\rm if}\quad \sigma = \sigma_0\cdot(t,\MRead{\iota})\cdot\sigma_1\ \&\
\sigma_0\reord{}({t},\MRead{\iota})\ {\rm or}
\\[3pt]&&
\phantom{{\rm if}\quad }\sigma_0=\delta_0\cdot(t',\Write{p,v}^{\Tid,I\cup\{\iota\}})\cdot
\delta_1\ \&\ \delta_0\reord{}(t',\Write{p,v}^{\Tid,I\cup\{\iota\}})
\\[3pt]
(S,\sigma, T) &\rlxm&
(S[p\dpeq v],\sigma_0\cdot\sigma_1, T)
&\kern18pt R4\ (\textit{write})
\\[3pt]&&
{\rm if}\quad \sigma = \sigma_0\cdot(t,\Write{p,v}^{W,I})\cdot\sigma_1\ \&\
\sigma_0\reord{}({t},\Write{p,v}^{W,I})\ \&\ v\in\Val
\\[5pt]
(S,\sigma, T) &\rlxm&
(S,\sigma_0\cdot(t,\Write{\varrho,v}^{W',I})\cdot\sigma_1, T)
&\kern-5pt{R5\ (\textit{write early})}
\\[3pt]&&
{\rm if}\quad \sigma=\sigma_0\cdot(t,\Write{\varrho,v}^{W,I})\cdot\sigma_1\ \&\
{t}\in W'\ \&\ W\subset W'\in\Visi
\\[3pt]
(S,\sigma, T) &\rlxm&
(S,\sigma_0\cdot\sigma_1, T)
&\kern11pt R6\ (\textit{barrier})
\\[3pt]&&
{\rm if}\quad \sigma = \sigma_0\cdot(t,\mo)\cdot\sigma_1\ \&\
\sigma_0\reord{}({t},\mo)
\end{array}
\\[-2pt]
\]
}{relaxopsemfigM}{$\mm$-Relaxed Operational Semantics (Memory)}

We now comment the rules. In all cases but the early ones ($R2$ and $R5$), 
performing an operation from 
the temporary store $\sigma$ consists in checking that the operation can be moved, up to 
$\reord{}$, at the head of $\sigma$, and then in removing the 
operation from $\sigma$ while possibly performing some effect. Namely, such an effect is 
produced when the performed operation is a read or a write. The 
reference that is concerned by the effect must be known in these cases. A read may also
return a value if it can be moved to a corresponding, 
visible write ($R2$). In this case, the read operation should 
not be blocked by barriers previously issued but not yet globally performed. 
This is expressed as $\sigma_0\reord{\Bar}({t},\Read{p,\iota})$.
The read operation does not completely vanish, but is transformed in a read mark
$\MRead{\iota}$, where $\iota$ identifies the matching write. 
When read is resolved using rule $R2$, 
the identifier of the read is added to the $I$ set of the write used to serve the read. 
The purpose of this set is to maintain the ordering of some memory operations,
as explained below in the subsection \ref{mmrequirements}.
As we shall see in Section \ref{globbarsect}, a read mark is only useful in relation 
with barriers and can be eliminated from the temporary store as specified by $R3$.
Notice that when we say that the read $(t,\Read{p,\iota})$ can be ``moved,'' this is
only an image: there is no transformation of the temporary store, but only a condition on
it, namely, in $R1$, $\sigma_0\reord{}(t,\Read{p,\iota})$. In the rules $R1$ and
$R2$ for read operations, there is a global replacement of the identifier $\iota$ associated
with the read by the actual value $v$ that is read: in these rules
$\substi{\iota}{v}(\sigma,T)$ stands for such a replacement, which does not affect the $I$
component in the writes. (Recall that we required
that an identifier such as $\iota$ cannot appear in the store.) Similarly, a write operation 
$\Write{\varrho,v}^{W,I}$ from the temporary store
$\sigma_0\cdot(t,\Write{\varrho,v}^{W,I})\cdot\sigma_1$ may update (rule $R4$) the memory
when $\varrho$ is a reference $p$, $v$ is in $\Val$ and the write is allowed to commute 
with the preceding operations, that is $\sigma_0\reord{}(t,\Write{p,v}^{W})$. An early 
write action in $R5$ has only the effect of modifying the temporary store, by extending 
the visibility of the write to more threads.

An obvious remark about the relaxed semantics is that it contains in a sense the 
interleaving semantics, with temporary stores containing at most one operation: 
one can mimick a transition of the latter either by one local
step, or by a local step immediately followed by a global action.
One can also immediately see that if $\Visi=\{\emptyset\}$, then the rule $R5$ cannot be
used, and consequently no early read can take place. If, in addition to $\emptyset$, 
$\Visi$ only contains the singletons $\{t\}$ for $t\in\Tid$, the read early rule $R2$ 
is restricted to the ``read-own-write-early'' capability \cite{AdGha}. 
In the write early rule $R5$, the 
requirement $t\in W'$ means that we do not consider memory models where the 
``read-others'-write-early'' capability would be enabled, but not the 
``read-own-write-early'' one (again, see \cite{AdGha}).

In the full version of the paper we also provide a formalization of speculative computations
in our framework.

\nssec{Memory Models: Requirements}\label{mmrequirements}
In the next section we briefly illustrate the expressive power of our framework for relaxed 
computations, by showing some programs exhibiting behaviors that are \textit{not} allowed by 
the reference semantics. (Many more examples are given in the full version of this paper.)
Most of these examples are
standard ``litmus tests'' found in the literature about memory models, that reveal in 
particular the consequences of relaxing in various ways the normal order of evaluation.
In most cases, the relaxations of program order can be specified by a binary relation
on $\tmp$. It is actually more convenient to use the converse relation, which can 
usually be more concisely described. We call this a \textit{precedence} relation.
Given such a binary relation $\Dep$ on pairs $(t,\xi)\in\tmp$, the commutability 
relation is supposed to satisfy
\[
(\omega,\xi)\,\Dep\,(\omega',\xi')\ \impl\ 
\forall\sigma,\sigma'.\ 
\neg\big(\sigma\cdot(\omega,\xi)\cdot\sigma'\reord{}(\omega',\xi')\big)
\]
That is, an operation in a temporary store is prevented from being globally performed
by another, previously issued one, that has precedence over it. A more positive 
formulation of this property is:
\begin{align}
\sigma\cdot(\omega,\xi)\cdot\sigma'\reord{}(\omega',\xi')
\ \impl\ \neg\big((\omega,\xi)\,\Dep\,(\omega',\xi')\big)
\tag{A${}_{\Dep}$}
\end{align}
Before examining various relaxations of the program order, by way of examples, 
we discuss some precedence pairs that 
are most often assumed in memory models. For instance, if we do not assume any constraint 
as regards the commutability of writes, from the program 
\[
(p\dpeq\true)\seq(p\dpeq\false)
\]
we could get as a possible outcome a state where the value of $p$ in the memory is $\true$,
by commuting the second write before the first.
This is clearly unacceptable, because this violates the semantics
of sequential programs. Then we should assume that two writes on the same reference issued
by the same thread cannot be permuted. Similarly, a write should not be overtaken by a read on 
the same reference issued by the same thread, and conversely, otherwise the semantics of
the sequential programs
\[\begin{array}{l}
(p\dpeq\true)\seq(r\dpeq\emi\,p)
\\[3pt]
(r\dpeq\emi\,p)\seq(p\dpeq\true)
\end{array}
\]
would be violated. We shall then require that any memory model satisfies axiom (A${}_{\ant}$)
where $\ant$ is the minimal precedence relation enjoying the following properties,
where the free symbols are implicitly universally quantified:
\[
\begin{array}{rcl}
\left.\begin{array}{r}
\varrho\in\{\varrho'\}\cup\Id\ \&
\\[7pt] t'\in W\cup\{t\}\ {\rm or}\ I\ne\emptyset\ne I'
\end{array}\right\} 
&\impl& 
\left\{\begin{array}{l}
(t,\Write{\varrho,v}^{W,I})\ant(t',\Read{\varrho',\iota})\ \&
\\[7pt]
(t,\Write{\varrho,v}^{W,I})\ant(t',\Write{\varrho',v'}^{W',I'})
\end{array}\right.
\\[15pt]
\iota\in I &\impl& (t,\Write{p,v}^{W,I})\ant(t',\MRead{\iota})
\\[3pt]
\varrho\in\{\varrho'\}\cup\Id &\impl&
(t,\Read{\varrho,\iota})\ant(t,\Write{\varrho',v}^{W,I})
\end{array}
\]
These properties ensure in particular that the precedence relations discussed 
above are enforced: among the operations of a given thread, one cannot commute for instance 
a read and a write on the same reference. 
Notice however that it is not required that the program order is
maintained as regards two reads on the same reference. 
Therefore, from the program
\\[-7pt]
\[
p\dpeq\true\ \para\ 
\begin{array}{l}
r_0\dpeq\emi p \seq
\\[2pt]
r_1\dpeq\emi p
\end{array}
\]
\\[-4pt]
if initially $S(p)=\false$,
we could end up in a state where the value of $r_1$ is $\false$, while the one for $r_0$
is $\true$. 
If one wishes to preclude such a behavior, one can simply add 
\[
\varrho\in\{p\}\cup\Id\ \impl\ (t,\Read{\varrho,\iota})\,\Dep\,(t,\Read{p,\iota'})
\]
to the precedence relation.

There are three cases where the $\ant$ precedence relates two distinct threads.
The first one, that is $(t,\Write{\varrho,v}^{W,I})\ant(t',\Read{p,\iota})$ where
$t'\in W$, means that a thread $t'$ ``sees'' the writes, previously issued 
by other threads, that include $t'$ in their scope -- the same holds with 
$(t,\Write{\varrho,v}^{W,I})\ant(t',\Write{\varrho',v'}^{W',I'})$ where $t'\in W$.
The precedence $(t,\Write{\varrho,v}^{W,I})\ant(t',\Write{\varrho',v'}^{W',I'})$ where 
$I\ne\emptyset\ne I'$ means that the order of writes on a given reference
must be respected if these writes have been read by some threads (this is similar to the 
``coherence order'' of \cite{SarkPLDI}). Finally, $(t,\Write{p,v}^{W,I})\ant(t',\MRead{\iota})$
where $\iota\in I$ means that an early read cannot vanish from the temporary store before 
the corresponding write. These two properties explains the role of the $I$ component in our model.
One should notice that \textit{no} specific precedence assumption is 
made at this point regarding the barriers.
Then our definition of the notion of a memory model is as follows:

\kl{memorymodeldef}{Definition (Memory Models)}{A {\rm memory model} $\mm$ for $\L$ 
is a pair $(\reord{},\Visi)$ where $\emptyset\in\Visi$, and the commutability 
predicate $\reord{}\ \sset\ \tmp^*\times\tmp$ satisfies the axioms {\rm(E)} and 
{\rm(A${}_{\ant}$)}.
}
As an example memory model, one can define $SC$, for $S$equen\-tial $C$onsistency, as 
\[
SC = (\{\varepsilon\}\times\tmp,\{\emptyset\}\cup\setof{\{t\}}{t\in\Tid}) 
\]
which obviously satisfies Definition~\memorymodeldef\ (the axiom (A${}_{\ant}$) is vacuously 
true). All the examples discussed in the following section 
hold in the minimal, or \textit{most relaxed}, memory model $\mm_{\ant}(\L)=(\reord{}_{\ant},
\powerset{\Tid})$, where $\reord{}_{\ant}$ is the largest commutability predicate 
satisfying (A${}_{\ant}$), $\powerset{\Tid}$ is the set of all subsets of $\Tid$.

In our work we mainly use commutability properties that are generated by precedence
relations, in the sense of axiom (A$_{\Dep}$). Then one could think of defining a memory 
model as a pair $(\Dep,\Visi)$, instead of $(\reord{},\Visi)$. However, we shall see
in Section~\ref{globbarsect} a case where this is not general enough.
More precisely, we
shall see a case where we have to say that $\neg\big(\sigma\reord{}(t,\xi)\big)$, not
on the basis that $\sigma$ contains an operation that has precedence over $(t,\xi)$,
but because there is a subsequence of $\sigma$ which, as a whole, 
has precedence over it.

\nsec{Examples}
Now we examine a few examples of programs exhibiting relaxed behaviors
that are {not} allowed by the reference semantics. (As mentioned above, in the full 
version of this paper we examine many more examples.)
In all the examples we assume that the initial values of the references are $\false$. 
We shall omit the superscript $W$ in $(t,\Write{\varrho,v}^{W,I})$ whenever $W=\emptyset$,
and similarly for $I$.

\nssec{Simple Relaxations}
Let us start with the most common relaxation, the one of the {\bf W$\arr$R}
order \cite{AdGha}, supported by simple write buffering as in TSO machines. 
That is, we are assuming that a write $(t,\Write{p,v}^{W,I})$ does not have
precedence over a read
$(t,\Read{q,\iota})$ if $p\ne q$. The litmus test here is the thread system $T$ of
the example (\ref{tsoex}) given in the Introduction (with an obvious thread names 
assignment). If we let 
\[
\sigma = (t_0,\Write{p,\true})\cdot(t_0,\Read{q,\iota_0})\cdot
(t_1,\Write{q,\true})\cdot(t_1,\Read{p,\iota_1})
\]
we have
\[
(S,\varepsilon,T) \xarr{\reord{},\Visi}{*}
(S,\sigma,(t_0,r_0\dpeq\iota_0)\para(t_1,r_1\dpeq\iota_1))
\]
It is then easy to see that, given that the order {\bf W$\arr$R} may be relaxed, we have
\[
(S,\varepsilon,T) \xarr{\reord{},\Visi}{*} (S,\sigma',
(t_0,r_0\dpeq\false)\para(t_1,r_1\dpeq\false))
\]
where $\sigma'=(t_0,\Write{p,\true})\cdot(t_1,\Write{q,\true})$. These
write operations can now be executed, and we reach a final state $(S',\varepsilon,T')$ 
where $S'(p)=\true=S'(q)$ and $S'(r_0)=\false=S'(r_1)$.

To restore $SC$ behavior in a relaxed memory model, the language must offer appropriate
synchronization means. Most often these are \textit{barriers},
that disallow some relaxations, when inserted between memory operations. For instance,
to forbid the {\bf W$\arr$R} relaxation, a natural barrier to use is $\wr$ (write/read), 
which cannot overtake a write, and cannot be overtaken by a read from the same thread. 
In our framework, the 
semantics of barriers are specified by the commutability predicate: they have no other
effect than preventing some reorderings. In the case of $\wr$, we require that the 
commutability predicate satisfies (A${}_{\Dep_{\wr}}$) for a precedence relation 
$\Dep_{\wr}$ such that
\[
(t,\Write{\varrho,v}^{W,I})\,\Dep_{\wr}\,(t,\wr)\,\Dep_{\wr}\,(t,\Read{p,\iota})
\]
(We do not have to specify that $\wr$ has precedence over $\MRead{\iota}$, because,
due to the conditions in $R2$, a read mark is never preceded by a read barrier in the 
temporary store.)
This is a \textit{local} barrier since it blocks only operations from the thread that 
issued it.
Then for restoring an $SC$ behavior to the example we are discussing, it is enough to 
insert this barrier in both threads:
\[
\begin{array}{l}
p\dpeq\true\seq 
\\
\wr\seq
\\
r_0\dpeq\emi\,q
\end{array}
\Big{\Vert}
\begin{array}{l}
q\dpeq\true\seq
\\
\wr\seq
\\
r_1\dpeq\emi\,p
\end{array}
\]
The threads will issue $\wr$ before the reads $\Read{q,\iota_0}$ and $\Read{p,\iota_1}$.
Given the precedence relations we just assumed as a semantics for $\wr$,
these reads cannot proceed until the barrier has disappeared from the temporary store.
The rule $R8$ requires, for a barrier to vanish, that it may be commuted with the 
previously issued operations. Then in the example above, this can only happen for
$\wr$ once the writes $\Write{p,\true}$ and $\Write{q,\true}$ have been globally 
performed.

We can deal in a similar way with the relaxation of the order {\bf W$\arr$W}, which when 
added to the previous relaxation characterizes the PSO memory model. And similarly with 
{\bf R$\arr$R} and {\bf R$\arr$W} which are sufficient to characterize the RMO model 
as described in \cite{AdGha}. In each case a corresponding local barrier, 
$\ww$, $\rr$ or $\rw$ can be used to restore sequential consistency. 

\nssec{Early Reads and Writes}\label{earlysect}

In this subsection, and the following one, we are concerned with architectures relaxing 
the atomicity of writes.
There are several examples to illustrate the write early rule $R5$, in combination
with $R2$, to show the ability for a thread to ``\textbf{read-own-write-early}''
or ``\textbf{read-others'-write-early}'', according to the terminology of 
\cite{AdGha}, that is the ability for a thread to read a write that has been previously 
issued, either by the thread itself or by a foreign thread, before the write updates
the shared memory. An example of the first, which holds in TSO models, is as follows:
\[\begin{array}{l}
p\dpeq\true\seq
\\
r_0\dpeq\emi\,p\seq\ (\true)
\\
r_1\dpeq\emi\,q\ (\false)
\end{array}
\Big{\Vert}
\begin{array}{l}
q\dpeq\true\seq
\\
r_2\dpeq\emi\,q\seq\ (\true)
\\
r_3\dpeq\emi\,p\ (\false)
\end{array}
\]
where the unexpected outcome is indicated by the annotations $(\true)$ and $(\false)$
associated with the assignments.
Let us assume that the write grain $\Visi$ contains two sets $W_0$ and $W_1$ such that
$t_0\in W_0$ and $t_1\in W_1$. Then it is easy to see that from this thread system we can, 
using the write early rule $R4$, reach a configuration where the temporary store is 
$\sigma_0\cdot\sigma_1$ where
\[
\begin{array}{rcl}
\sigma_0 &=& (t_0,\Write{p,\true}^{W_0})\cdot(t_0,\Read{p,\iota_0})\cdot
(t_0,\Write{r_0,\iota_0})\cdot(t_0,\Read{q,\iota_1})
\\[3pt]
\sigma_1 &=& (t_1,\Write{q,\true}^{W_1})\cdot(t_1,\Read{q,\iota_2})\cdot
(t_1,\Write{r_2,\iota_2})\cdot(t_1,\Read{p,\iota_3})
\end{array}
\]
Then by $R2$ both $\iota_0$ and $\iota_2$ can take the value $\true$, whereas, given
that the order {\bf W$\arr$R} is relaxed (and that a read mark does not have precedence
over a read), both $\iota_1$ and $\iota_3$ take the value
$\false$ from the shared store, before it is updated by performing the writes
$\Write{p,\true}^{W_0}$ and $\Write{q,\true}^{W_1}$.

As regards the \textbf{read-others-write-early} ability, the best known litmus test is IRIW 
(Independent Reads of Independent Writes):
\[
p\dpeq\true\ \big{\Vert}\ q\dpeq\true\ \big{\Vert}\
\begin{array}{l}
r_0\dpeq\emi\,p\seq\ (\true)
\\
r_1\dpeq\emi\,q\ (\false)
\end{array}
\ \big{\Vert}\
\begin{array}{l}
r_2\dpeq\emi\,q\seq\ (\true)
\\
r_3\dpeq\emi\,p\ (\false)
\end{array}
\]
In our framework, this example is accounted for in the following way. 
Assume that $\Visi$ contains two sets $W_0$ and $W_1$ such that $\{t_0,t_2\}\sset W_0$ and
$\{t_1,t_3\}\sset W_1$, with $t_3\not\in W_0$ and $t_2\not\in W_1$. Then we have, using $R5$ 
twice:
\[
(S,\varepsilon,T) \xarr{\reord{},\Visi}{*}
(S,(t_0,\Write{p,\true}^{W_0,\emptyset})\cdot(t_2,\Read{p,\iota_0})\cdot
(t_1,\Write{q,\true}^{W_1,\emptyset})\cdot(t_3,\Read{q,\iota_2}),T')
\]
Now since the write of $p$ is made visible to thread $t_2$, the identifier $\iota_0$ can
take the value $\true$, and similarly $\iota_2$ takes the value $\true$, by the rule $R2$. 
Since the writes from $t_0$ and $t_1$ are not visible from $t_3$ and $t_2$ respectively,
these threads may read the value $\false$ from the shared memory for both $q$ and $p$.
One finally reaches a state where $S'(r_0)=\true=S'(r_2)$ whereas $S'(r_1)=\false=S'(r_3)$. 
Notice that in this computation
we never have to ``commute'' operations (the precedence relation could be anything here), 
that is, this computation proceeds in program order, and therefore inserting local 
barriers in $t_2$ and $t_3$ would not influence it. 
Similar examples that are discussed in \cite{BoAd,SarkPLDI}, such as  WRC, RWC and CC, 
can be explained in the same way. This is the case for instance of WRC (Write-to-Read 
Causality) -- without $\mathsf{fence}$ since, as with IRIW, we follow the program order 
here:
\[
p\dpeq\true\ \big{\Vert}\ 
\begin{array}{l}
r_0\dpeq\emi\,p\seq\ {(\true)}
\\[2pt]
q\dpeq\true
\end{array}
\big{\Vert}
\begin{array}{l}
r_1\dpeq\emi\,q\seq\ {(\true)}
\\[2pt]
r_2\dpeq\emi\,p\ {(\false)}
\end{array}
\]
Here the write $(p\dpeq\true)$ is issued, and, with some appropriate assumption about the
write grain, made visible to the second thread (but not to the third), which
will then assign the value $\true$ to $r_0$. Then the write $(q\dpeq\true)$ is globally
performed, and, before the operation $\Write{p,\true}^{W,I}$ reaches the store,
the third thread is executed, reading the values $\true$ for $q$ in $(r_1\dpeq\emi\,q)$
and $\false$ for $p$ in $(r_2\dpeq\emi\,p)$. That is, the  outcome $S'(r_0)=\true=S'(r_1)$ 
and $S'(r_2)=\false$ is allowed. 

\nssec{Global Barriers}\label{globbarsect}
In a model that enables the read-others'-write-early capability, one needs in the language some 
barrier having a \textit{global} effect on writes, that is, a barrier that is prevented from 
vanishing by writes from foreign threads. We shall use here the case of PowerPC, as 
described by \cite{SarkPLDI}, to exemplify the framework. Indeed,
the PowerPC architecture offers such a strong $\sync$ barrier, which imposes the 
program order to be preserved between any pair of (local) reads and writes. 
This means that it enjoys the same precedence relations as $\wr$, $\ww$, $\rr$ and $\rw$.
The global effect of $\sync$ is the one suggested above: $\sync$ maintains the order
between two writes, the first one being a
visible write from a foreign thread, and the second being a local write. Then to specify
the semantics of this barrier we just have to add the following:
\[
t'\in W\ \impl\ (t,\Write{\varrho,v}^{W,I})\,\Dep_{\sync}\,(t',\sync)
\]
The PowerPC architecture also provides an $\lwsync$ barrier, which is weaker than
$\sync$. First, this is a $\ww$, $\rw$ and $\rr$ barrier, but it does not order the pairs
of writes and reads, to preserve some TSO optimizations. Therefore, we cannot define 
the semantics of $\lwsync$ by means of a binary precedence relation, as we did up to now.
Nevertheless, the following precedences are part of the semantics of $\lwsync$ in
our framework:
\[ 
\begin{array}{c}
(t,\MRead{\iota})\,\Dep_{\lw}\,(t,\lwsync)
\quad\&\quad
(t,\Read{\varrho,\iota})\,\Dep_{\lw}\,(t,\lwsync)\,\Dep_{\lw}\,(t,\Write{\varrho',v}^{W,I})
\\[3pt]
t=t'\ {\rm or}\ t'\in W\ \impl\ 
(t,\Write{\varrho,v}^{W,I})\,\Dep_{\lw}\,(t',\lwsync)
\end{array}
\]
Next, we have to say that $\lwsync$ is a $\rr$ barrier, even though it does not have 
precedence over reads. Then we assume that the commutability predicate satisfies the 
following:
\[
\left.\begin{array}{l}
\sigma = \sigma_0\cdot(t,\lwsync)\cdot\sigma_1\ \&\
\\[6pt]
\sigma_0 = \delta_0\cdot(t,\Read{\varrho,v})\cdot\delta_1\ {\rm or}\ 
\sigma_0 = \delta_0\cdot(t,\MRead{\iota})\cdot\delta_1
\end{array}\right\}
\ \impl\ \neg\big(\sigma\reord{\lw}(t,\Read{p,v'})\big)
\]
This completes the definition of the semantics of $\lwsync$. 
Let us see two examples. If we insert global barriers into the IRIW configuration,
as follows:
\[
p\dpeq\true\ \big{\Vert}\ q\dpeq\true\ \big{\Vert}\
\begin{array}{l}
r_0\dpeq\emi\,p\seq\ (\true)
\\
\lwsync\seq
\\
r_1\dpeq\emi\,q\ (\false)
\end{array}
\big{\Vert}
\begin{array}{l}
r_2\dpeq\emi\,q\seq\ (\true)
\\
\sync\seq
\\
r_3\dpeq\emi\,p\ (\false)
\end{array}
\]
then the unexpected outcome is still not prevented to occur. This is obtained
as follows: the operation of the second thread ($t_1$) is issued, and then the ones
of $t_3$, $t_0$ and $t_2$, in that order. Then
the visibility of $(t_0,\Write{p,\true})$ is made global, and therefore $t_2$ can
read the value $\true$ for $p$. 
Since the write from $t_0$ is allowed to be performed immediately, the read mark
left when performing $\Read{p,\iota_0}$ may disappear. 
The $\lwsync$ from $t_2$ is still prevented to vanish by the write from
$t_0$, but it no longer blocks the second read of $t_2$.

In the case of the WRC litmus test \cite{BoAd}, inserting $\lwsync$ barriers
prevents the unexpected outcome showed in
\[
p\dpeq\true
\ \Big{\Vert}\ 
\begin{array}{l}
r_0\dpeq\emi\,p\seq\ (\true)
\\
\lwsync\seq
\\
q\dpeq\true
\end{array}
\Big{\Vert}\ 
\begin{array}{l}
r_1\dpeq\emi\,q\seq\ (\true)
\\
\lwsync\seq
\\
r_2\dpeq\emi\,p\ (\false)
\end{array}
\]
to occur. Similarly, inserting $\sync$ barriers in the third and fourth threads of
the IRIW example restores an SC behavior.
To see this, we have to explore all the possible behaviors, and this is where our
software tool is useful.

\nsec{The Simulator}
The set of configurations that may be reached by running a program in the relaxed 
semantics can be fairly large, and it is sometimes difficult, and error prone, 
to find a path to some (un)expected final state, or to convince 
oneself that such an outcome is actually forbidden, that is, unreachable. 
Then, to experiment with our framework, 
we found it useful to design and implement a simulator that allows us to exhaustively
explore all the possible relaxed behaviors of (simple) programs.
As usual, we have to face a state explosion problem, which is much worse than with
the standard interleaving semantics.

Our simulator is written in \textsf{J{\small AVA}}. Its main function \verb|step|
computes all the configurations reachable in one step from a given configuration.
A brute force simulator would then recursively use the \verb|step| function, in a depth 
first manner, in order to compute reachable configurations that have an empty temporary store 
and a terminated thread pool, where all the thread expressions are values. This methodology 
does not consume much memory space, being basically proportional
to the \verb|log| of the number of reachable states or, similarly, to the depth
of the tree induced by the \verb|step| function. However, the number of
configurations in this tree grows very fast with the size of the expression to analyse.
For instance, with the example (\ref{tsoex}) given in the Introduction, this
brute force strategy has been aborted after generating more than
$20\times 10^{10}$ configurations and after half a day of computing, even
if it is obvious that only four {\em different} final configurations may be reached. 
Therefore, a first improvement is to transform the tree
traversal by a dag construction  merging all the same configurations. Less configurations
will be constructed and analyzed (only $60\,588$ for the example), but all these configurations
must be simultaneously in memory.

Several other optimizations have been used.
In order to reduce the search space, in the simulator we use a refined rule $R5$ where 
the visibility set $W'$ is supposed to be either $\Tid$ or a subset of $\live(T)\cup
\rdt(\sigma_1)$ where the sets $\live(T)$ and $\rdt(\sigma)$ of thread identifiers are 
defined as follows:
\[
\begin{array}{rcl}
\live(\emptyset) &=& \emptyset
\\[2pt]
\live((t,e)\para T) &=& \live(T)\cup\setof{t}{e\not\in\Val}
\end{array}
\quad
\begin{array}{rcl}
\rdt(\varepsilon) &=& \emptyset
\\[2pt]
\rdt((t,\xi)\cdot\sigma) &=& \rdt(\sigma)\cup\setof{t}{\exists\varrho,\iota.\ 
\xi=\Read{\varrho,\iota}}
\end{array}
\]
We have not presented this formulation in Figure~\relaxopsemfigM\ only because it is 
conceptually a bit more obscure. With this optimization, in our example, the
number of configurations falls down from $60\,588$ to $51\,068$. 
A more dramatic optimization is obtained by introducing a distinction between ``registers,''
that are local to some thread, and shared references. As suggested above, the registers 
are denoted $r_i$ in the examples. Indeed, these registers are not concerned by
early reads from foreign threads, and therefore applications of the rule $R5$ to them
may be drastically restricted. In this way,
the number of generated configurations in the case of example (\ref{tsoex}) decreases
from $51\,068$ to $13\,356$ for instance.
Furthermore, one may observe that, since removing an operation from a temporary store
$\sigma$ never depends on what follows this operation in $\sigma$, the strategy that
consists in applying first the rules of Figure~\relaxopsemfigT\ for evaluating the threads
before attempting anything else (that is, applying a rule from Figure~\relaxopsemfigM)
will never miss any final configuration. This allows us to generate only $2\,814$ 
configurations in the case of example (\ref{tsoex}) for instance.

However, the optimized search strategy outlined above still fails in exploring exhaustively 
some complex litmus tests.
In such cases, we make a tradeoff between time and space: for each temporary store that
can be reached by applying the rules of Figure~\relaxopsemfigT\ as far as possible, we 
generate the reachable final configurations, but we do not share this state space 
among the various possible temporary stores. For instance, still regarding the example
(\ref{tsoex}), there are $20$ possible ``maximal'' temporary stores, and running 
independently the simulator in each case generates an average number of $500$ configurations,
so that the total of number of generated configurations following this simulation method
raises up to $10\,280$.
Nevertheless this allowed us to successfully explore a large number of litmus tests,
and in particular all the ones presented
by Sarkar \& al.\ \cite{SarkPLDI} in their web files. We report upon this in the full 
version of the paper. Our simulator is available on the web page
\texttt{http://www-sop.inria.fr/indes/MemoryModels/}.

\nsec{Conclusion}
We have introduced a new, operational way to formalize the relaxed semantics of concurrent
programs. Our model is flexible enough to account for a wide variety of weak behaviors,
and in particular the odd ones occurring in a memory model that does not preserve the
atomicity of writes. To our view, our model is also simple enough to be easily understood 
by the implementer and the programmer, and precise enough to be used in the formal
analysis of programs.

\providecommand{\urlalt}[2]{\href{#1}{#2}}
\providecommand{\doi}[1]{doi:\urlalt{http://dx.doi.org/#1}{#1}}

\def\LNCS{Lecture Notes in Comput.\ Sci.\ }

\def\kref#1#2#3#4#5#6#7{\bibitem{#1}
\textsc{#2, }\textsl{#3, }#4 ({\oldstyle #5}) #6. \doi{#7}}
\def\krapp#1#2#3#4#5{\bibitem{#1}
\textsc{#2, }\textsl{#3, }#4 ({\oldstyle #5}).}
\def\p{\kern -1pt{.}\kern 2pt}
\def\pp{\kern -1pt{.}\kern 0.7pt}

\def\AdveBoehmCACM#1{\kref{#1}{S\p Adve, H.-J\p Boehm}{Memory models:
a case for rethinking parallel languages and hardware}{CACM
Vol.\ 53 No.\ 8}{2010}{90-101}{10.1145/1787234.1787255}}

\def\AdveGharachorlooIEEE#1{\kref{#1}{S\p Adve, K\p Gharachorloo}
{Shared memory consistency models: a tutorial}{IEEE Computer Vol.\ 29
No.\ 12}{1996}{66-76}{10.1109/2.546611}}

\def\AdveHillISCA#1{\kref{#1}{S\p Adve, M\pp D\p Hill}{Weak ordering -- 
A new definition}{ISCA'90}{1990}{2-14}{10.1145/325096.325100}}

\def\AtigESOP#1{\kref{#1}{M\p Atig, 
A\p Bouajjani, S\p Burckhardt, M\p Musuvathi}{What's Decidable 
about Weak Memory Models?}{ESOP'12}{2012}{26-46}{10.1007/978-3-642-28869-2\_2}}

\def\BattyEtAlPOPL#1{\kref{#1}{M\p Batty, S\p Owens, S\p Sarkar, 
P\p Sewell, T\p Weber}{Mathematizing C++ concurrency}{POPL'11}{2011}
{55-66}{10.1145/1925844.1926394}}

\def\BoehmAdvePLDI08#1{\kref{#1}{H.-J\p Boehm, S\p Adve}{Foundations
of the C++ concurrency model}{PLDI'08}{2008}{68-78}{10.1145/1375581.1375591}}

\def\BoudolPetriMM#1{\kref{#1}{G\p Boudol, G\p Petri}{Relaxed memory models:
an operational approach}{POPL'09}{2009}{392-403}{10.1145/1480881.1480930}}

\def\BoudolPetriSpecESOP#1{\kref{#1}{G\p Boudol, G\p Petri}{A theory
of speculative computations}{ESOP'10, \LNCS 6012}{2010}{165-184}{10.1007/978-3-642-11957-6\_10}}

\def\BurckhardtMusuvathiSinghCC#1{\kref{#1}{S\p Burckhardt, M\p Musuvathi,
V\p Singh}{Verifying local transformations on relaxed memory models}
{CC'10, \LNCS 6011}{2010}{104-123}{10.1007/978-3-642-11970-5\_7}}

\def\GharachorlooISCA90#1{\kref{#1}{K\p Gharachorloo, D\p Lenoski, 
J\p Laudon, P\p Gibbons, A\p Gupta, J\p Hennessy}{Memory consistency
and event ordering in scalable shared-memory multiprocessors}{ACM SIGARCH
Computer Architecture News Vol.\ 18 No. 3a}{1990}{15-26}{10.1145/325164.325102}}

\def\LamportIEEETransComput#1{\kref{#1}{L\p Lamport}{How to make a
multiprocessor computer that correctly executes multiprocess programs}
{IEEE Trans.\ on Computers Vol.\ 28 No.\ 9}{1979}{690-691}{10.1109/TC.1979.1675439}}

\def\MansonPughAdvePOPL#1{\kref{#1}{J\p Manson, W\p Pugh, S\pp A\p Adve}
{The Java memory model}{POPL'05}{2005}{378-391}{10.1145/1040305.1040336}}

\def\ParkDillTOC#1{\kref{#1}{S\p Park, D\pp L\p Dill}{An executable
specification and verifier for Relaxed Memory Order}{IEEE Trans.\
on Computers, Vol.\ 48, No.\ 2}{1999}{227-235}{10.1109/12.752664}}

\def\SaraswatPPoPP#1{\kref{#1}{V\p Saraswat, R\p Jagadeesan, M\p Michael,
C\p \hbox{\rm von} Praun}{A theory of memory models}{PPoPP'07}{2007}
{161-172}{10.1145/1229428.1229469}}

\def\SarkarPLDI11#1{\kref{#1}{S\p Sarkar, P\p Sewell,  J\p Alglave,
L\p Maranget, D\p Williams}{Understanding POWER multiprocessors}
{PLDI'11}{2011}{175-186}{10.1145/1993498.1993520}}

\def\SewellEtAlX86TSO#1{\kref{#1}{P\p Sewell, S\p Sarkar, S\p Owens, 
F\p Zappa Nardelli, M\p O\p Myreen}{x86-TSO: A rigorous and usable 
programmer's model for x86 multiprocessors}{CACM Vol.\ 53 No.\ 7}
{2010}{89-97}{10.1145/1785414.1785443}}

\end{document}